\documentclass[%
superscriptaddress,
 reprint,
 amsmath,amssymb,
 aps,
floatfix,
]{revtex4-2}

\usepackage{graphicx}% Include figure files
\usepackage{dcolumn}% Align table columns on decimal point
\usepackage{bm}% bold math
\usepackage{hyperref}% add hypertext capabilities
\usepackage{comment}
\usepackage{csquotes}
\usepackage{braket}
\usepackage[toc,page]{appendix}
\begin{document}

\title{Structural inequalities exacerbate infection disparities: A computational approach}% Force line breaks with \\
%\thanks{A footnote to the article title}%

\author{Sina Sajjadi}
\email{sajjadi@csh.ac.at}
\affiliation{Complexity Science Hub Vienna, Vienna, Austria}
\affiliation{Central European University, Vienna, Austria}

 %\altaffiliation[Also at ]{Physics Department, XYZ University.}%Lines break automatically or can be forced with \\
\author{Pourya Toranj Simin}%
\email{pourya.toranj-simin@inserm.fr}
\affiliation{
Shahid Beheshti University, Tehran, Iran}%

%\collaboration{MUSO Collaboration}%\noaffiliation

\author{Mehrzad Shadmangohar}
\affiliation{
Shahid Beheshti University, Tehran, Iran}%
\author{Basak Taraktas}
\affiliation{%
 Bogazici University, Istanbul, Turkey
}%

\author{Ulya Bayram}
\affiliation{%
 Çanakkale Onsekiz Mart University, Çanakkale, Turkey
}%

\author{Maria V. Ruiz-Blondet}
\affiliation{%
Neurable, Boston, United States
}%

\author{Fariba Karimi}
\email{karimi@csh.ac.at}
\affiliation{Complexity Science Hub Vienna, Vienna, Austria}

%\collaboration{CLEO Collaboration}%\noaffiliation

%\date{\today}% It is always \today, today,
             %  but any date may be explicitly specified

\begin{abstract}
\textit{\textit{Background}}: During the COVID-19 pandemic, we witnessed a disproportionate infection rate among marginalized and low-income groups. Despite empirical evidence suggesting that structural inequalities in society contribute to health disparities, there has been little attempt to offer a computational and theoretical explanation to establish its plausibility and quantitative impact.
\textit{Methods}: This paper combines an agent-based network model and a compartmental susceptible-infectious-removed model (SIR) to explain how socioeconomic inequalities affect the dynamics of disease spreading. Specifically, we focus on two aspects of structural inequalities: wealth inequality and social segregation.
\textit{Findings}: Our computational model demonstrates that under high income inequality, the infection gap widens between the low-income and high-income groups, and also the overall infected cases increase.
We also observed that social segregation between different socioeconomic status (SES) groups intensifies the spreading and hence mortality rates.
Furthermore, we explain the second peak of the infectious cases during a pandemic as a result of a false sense of safety and loosening the quarantine, among the higher SES individuals.
\textit{Interpretation}: These findings send a strong message to policymakers; confinement measures must be accompanied by substantial financial assistance to those from lower-income groups so that the people regardless of their socioeconomic status can afford to stay in. Without financial assistance, lower-income individuals will remain in circulation, which will prolong the duration and magnitude of the infection.
\end{abstract}

%\keywords{Suggested keywords}%Use showkeys class option if keyword
                              %display desired
\maketitle

%\tableofcontents

\section{Introduction}

The COVID-19 pandemic resulted in a disproportionate infection rate among marginalized groups, people of color \cite{bambra2020covid}, and low-income groups \cite{adam2022pandemic}. Health inequalities are often due to existing structural inequalities in which societies foster discrimination through mutually reinforcing systems of housing, education, employment, and health care \cite{bailey2017structural}. It has long been established that pandemics expose and deepen existing racial, ethnic, and income inequalities in society by aggravating resource constraints and rendering living conditions direr for those who live on the margins \cite{Wade:2020aa}. Despite growing interest in understanding how social and structural factors drive inequalities in health outcomes, systematic and quantitative understanding of their effects as a root cause of health inequalities has not received much attention.
Inspired by the recent COVID-19 outbreak, this paper studies the effect of structural inequalities, namely income inequality and social segregation, on the infection disparities among different socioeconomic groups during public health crises. We propose a novel explanatory framework that incorporates game theory, agent-based modeling, and network analysis, to shed light on the impact of individual decision-making dynamics on the macro-level epidemic outcomes. 

More precisely, two crucial factors that contribute to structural inequalities, and subsequently, health inequalities are explored \cite{gee2011structural}: income inequalities, and social segregation. Income inequalities often manifested in socioeconomic status and correlated with race and ethnicity, influence multiple aspects of human life and decision-making processes. Research suggests that income inequality and marginalization increase infection rates among disadvantaged groups as a result of the inability to reduce mobility \cite{chang2021mobility}.

Social segregation is often the outcome of homophilous interactions \cite{mcpherson2001birds} along the racial \cite{Schelling, Black-White}, ethnicity \cite{SegregationFactors}, or occupational lines \cite{SegregationIndex}. This social tendency in the micro-level influences the social network structure in the macro-level \cite{karimi2018homophily} and impacts the spreading dynamics \cite{pei2013spreading,  rizi2021epidemic, hiraoka2021herd}. 

Evidence suggests that income inequalities and social segregation generate persistent long-term health inequalities between upper- and lower-income households \cite{InequlityInEU, SESInVaccination}, which we must take into account to make realistic predictions on epidemic diffusion patterns.

During public health crises, social distancing and confinement measures become of utmost priority to contain the spread of the infection. However, the level of compliance with these measures is closely related to the nature of working and living conditions. In addition, lower-income groups experience greater income loss as a result of the economic slowdown that lockdown and social distancing measures generate, which, in turn, may depress these groups' willingness to stay in and wait for the pandemic out. Thus it is important to consider the interplay between epidemic growth, income status, and the individuals' willingness to self-isolate over time. 
Existing empirical studies tend to focus on structural determinants of health inequality \cite{Williams:2010aa, Health-Statistics-U.-S.-:2012aa, linkphelan}, and long-term health implications of such inequality \cite{deathcolor, Faust:2020aa}. The few studies that addressed the decision-making dynamics (such as vaccination games) did not elaborate on the segregation aspect \cite{arefin2019interplay, AMARAL2021110616, CHEN2020125428}.
To this end, we study the effect of socioeconomic disparities and social segregation on infection and subsequently mortality rates using epidemiological models.

We compare the behavior of higher-income groups to the lower-income groups when governments enforce lockdown measures. 
This modeling framework overcomes multiple shortcomings in the previous simulation models of epidemics. First, many infection models utilize ordinary differential equations (ODEs) and assume that each infected individual infects every susceptible individual with an equal probability \cite{kermack1927contribution,fenichel2011adaptive,DeterministicSIR}.
Furthermore, while an extensive body of research has been focused on the dynamics of the underlying social contacts \cite{masuda2017temporal, gauvin2020randomized, rocha2013bursts, PhysRevE.104.014313, sajjadi2021impact}, the majority of the works consider the contact dynamics as an independent process, and ignore the infection-induced \cite{gross2006epidemic} social dynamics. Hence, we devise an adaptive network framework, allowing individuals to quarantine and temporarily inactivate their contacts, based on their income loss and infection exposure in their social network.   

In addition, existing computational studies of epidemics, including COVID-19 cases, use SIR models to calculate the efficiency \cite{sharov2020creating} and practicality of herd immunity strategies \cite{brett2020transmission}, or fractional compartmental models to make predictions of real-world scenarios \cite{hoan2020new, taghvaei2020fractional}. However, these studies do not consider the feasibility of public health strategies among individuals with varying levels of adaptability to the new policies. They also ignore social segregation as one of the drivers of disparity in infection cases among different socioeconomic groups. In reality, not only are individuals more likely to interact with those in their vicinity but also their level of compliance with lockdown measures vary by their socioeconomic status.
Finally, game theoretical models are effective in modeling trade-offs between optimal vaccination strategies for the community and individuals \cite{fine1986individual}, those between infection risk and economic cost \cite{kabir2020evolutionary}, as well as the effectiveness of social distancing \cite{reluga2010game} and individual decisions to comply with lockdown measures \cite{fenichel2011adaptive}. However, these models often do not incorporate social segregation in investigating diffusion dynamics \cite{SILVA2020110088}. Here we combine the strengths of these approaches, by embedding an adaptive game-theoretical model (to capture decision making with trade-offs) into an agent-based network model (that examines the effect of network variations in diffusion).
Our computational analysis confirms that the adverse effect of income inequality and social segregation amplify during health crises especially when governments enforce confinement measures to contain the spread of infection.

\section{Methods}
We model the spreading of an infection in a segregated society, where people (agents) need to decide to quarantine themselves or not as an adaptive dynamical process. We classify the population into $B$ socioeconomic (SES) groups/blocks,
where agents of higher SES can more easily afford to quarantine themselves.
Group membership also controls the interaction patterns of the agents, as there is a higher interaction probability \textit{within} the groups, compared to the interaction probability \textit{between} different groups.
Thus, our model consists of three components: a) contact network, b) decision-making process, and c) epidemic spreading. The components are described in the following:
\subsection*{Contact network}
We use a Stochastic Block Model (SBM) to model the segregation between different communities.
Every two agents $i$ and $j$ respectively belonging to socio-economic groups/blocks $a$ and $b$ are connected with a probability $\rho_{a,b}$, where $\boldsymbol{ \rho }$ is known as the connection probability matrix. Due to the residential segregation, the probability of interaction among individuals of the same group is higher than average.
Parameter $s$ controls the level of segregation, with $s=0$ and $s=1$, leading to a fully homogeneous and fully segregated communities respectively (see Fig. \ref{fig:s-network} in the Supplementary Information).
\subsection*{Decision making}
At each time step, every agent decides whether to \textit{quarantine} itself or to \textit{participate} in the society, i.e., go to work.
Following \cite{blume1993statistical, szabo1998evolutionary, hauert2005game, traulsen2007pairwise}
we use the Fermi decision making function to model the decision making process, 
where agents assign probability $e ^ {\beta r_{\omega}} / C$ to each option $\omega$, based on its possible reward/punishment $r$. $C$ serves as the normalization factor and $\beta$ indicates the intensity of selection. For $\beta \rightarrow \infty$ agents will almost definitely pick the option with the slightest reward advantage, while for $\beta \rightarrow 0$ they randomly choose; disregarding the rewards. For smaller positive $\beta$ values, agents assign a higher probability to the option with the higher reward.
Participation or going to work, exposes the agent to the infection, resulting in a \textit{psychological fear of infection punishment}.
The option of quarantining, in contrast, leads to \textit{income loss}.
We model the punishment of participation as $r_d I(t)$. Where $r_d$ denotes the baseline fear of infection and $I(t)$ is the total fraction of infectious agents at time $t$. Hence, the perceived risk of the infection is linearly dependent on the number of infectious agents in the network.
On the other hand, $r_b$ indicates the income-loss punishment for agents in block $b$.
The effect of income loss is conceived to be higher for individuals from lower socio-economic classes, as they will be less likely to afford to lose their income and quarantine themselves.
Hence, we hypothesize the punishment of income-loss to be inversely proportional to the wealth of each individual ($r_b = -\frac{1}{w}$).
Wealth $w$ is distributed by the Pareto distribution $p(w) =  \frac{\lambda w_m^\lambda}{w^{\lambda+1}}$ \cite{newman2005power, persky1992retrospectives}.
The parameter $\lambda$ controls the level of equality, where $\lambda \rightarrow 0$ and $\lambda \rightarrow \infty$ represent the ultimate inequality (Dirac delta distribution of wealth) and equality (Uniform distribution of wealth) respectively. Distribution of wealth in turn controls the fear of income-loss ($r_b = -\frac{1}{w}$)
(\textit{See section \ref{sec:economic-inequality-appendix} in the Supplementary Information.})
We thus, derive the probability $P_b$ of participation for agents of block $b$ to be:
\begin{equation}\label{eq:Prob-main}
P_b(t) = \frac{1}{e^{\beta \left( r_b - r_d I(t)\right)}  + 1 }
\end{equation}

\subsection*{Epidemic Spreading}
We use a SIR model to simulate the spread of an infectious disease in the network. We categorize the population into three compartments, $S$ (Susceptible), $I$ (Infectious), and $R$ (Removed).
It should be noted that the removed component, $R$, includes both the recovered and the deceased individuals, as both are incapable of further interacting with the disease. The proportion of the deceased individuals within the removed population can be subsequently calculated based on the specific disease, and need not be explicitly considered in the model.

Initially a fraction $I_{init}$ of agents in each block are in the state $I$, with the rest being in the state $S$.
Infection only propagates among the agents out of quarantine. At each time step, infectious agents $I$ infect their $S$ neighbors with probability $\mu \Delta t$, where $\mu$ is the transmission rate and $\Delta t$ is the size of a time step.
Infectious agents are removed with the removal probability $\gamma \Delta t$, where $\gamma$ is the removal rate, and they will not be able to get infected or infect others again.

\begin{figure*}[!ht]
	\begin{center}
		\includegraphics[width=\linewidth]{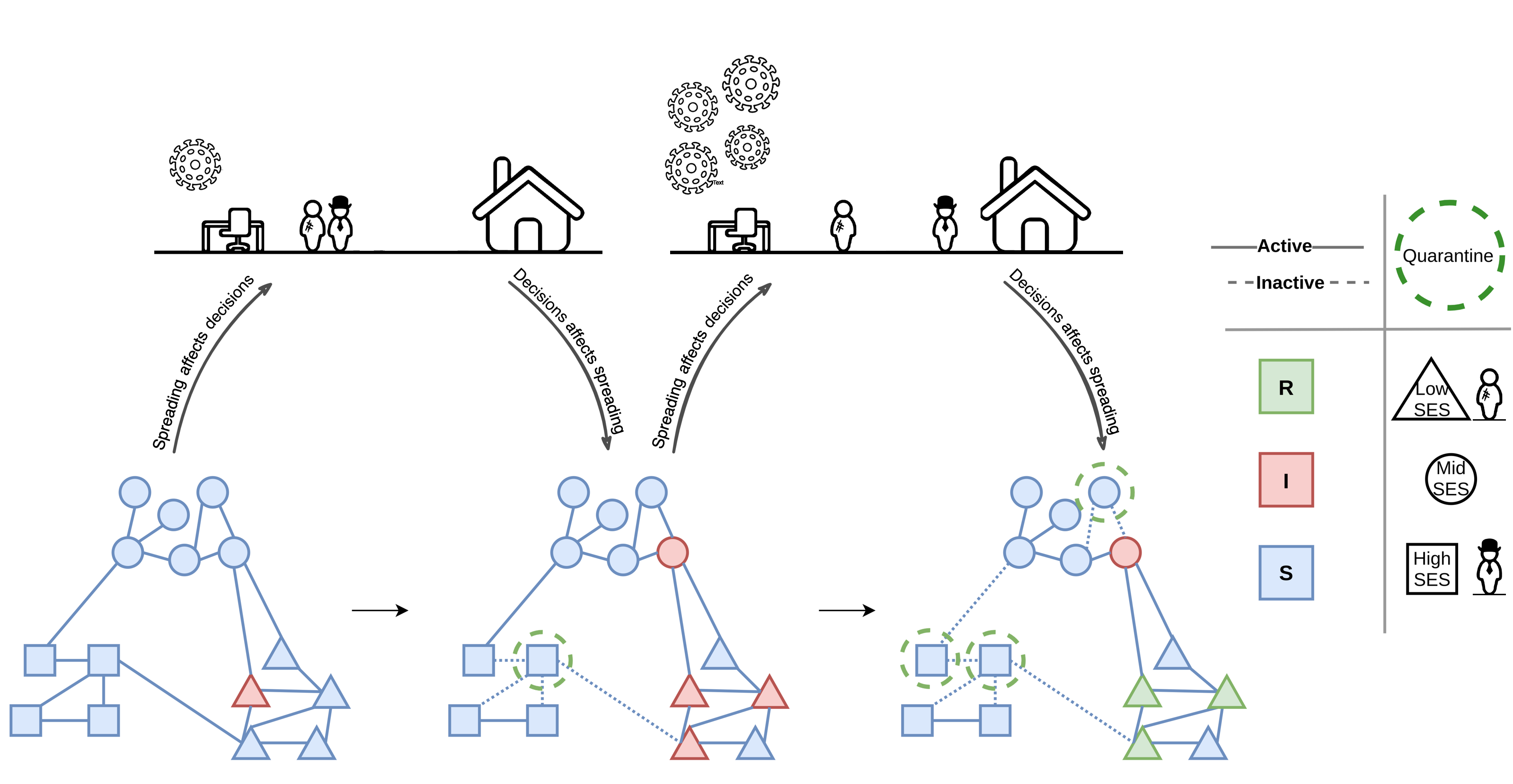}
		\caption{
		Schematic illustration of the interplay between the spreading and the decision-making models.
		On the top, we have shown the tendency of high and low SES individuals toward \textit{participation} (Office) and \textit{quarantine} (Home) under low and high infection scenarios.
		At the bottom we have shown the progress of the dynamics in three snapshots: Infectious agents transmit the disease to their neighbors. Some agents respond with quarantining (dashed green circle) and temporarily detaching themselves from their neighbors, eliminating the transmission possibility. Infectious agents get removed after some time. Shapes denote SES class/block and colors indicate infection status. Solid and dashed lines respectively illustrate the active links, capable of transmitting the disease, and inactive links, incapable of doing so.
		The curved arrows denote the interplay between the two dynamics resulting in the adaptive dynamics of our model.}
		\label{fig:model-illus}
	\end{center}
\end{figure*}

In Fig. \ref{fig:model-illus}, we illustrate a sample progress of the dynamics, where the infection spreads and agents respond with quarantining. We also devise a mean-field approximation (MFA) to further investigate the model mathematically and evaluate its robustness.
(\textit{See section \ref{sec:mean-field-derivation} in the Supplementary Information for the detailed derivation of the mean-field approximation.})

\begin{figure}[!ht]
	\begin{center}
	\includegraphics[width=\linewidth]{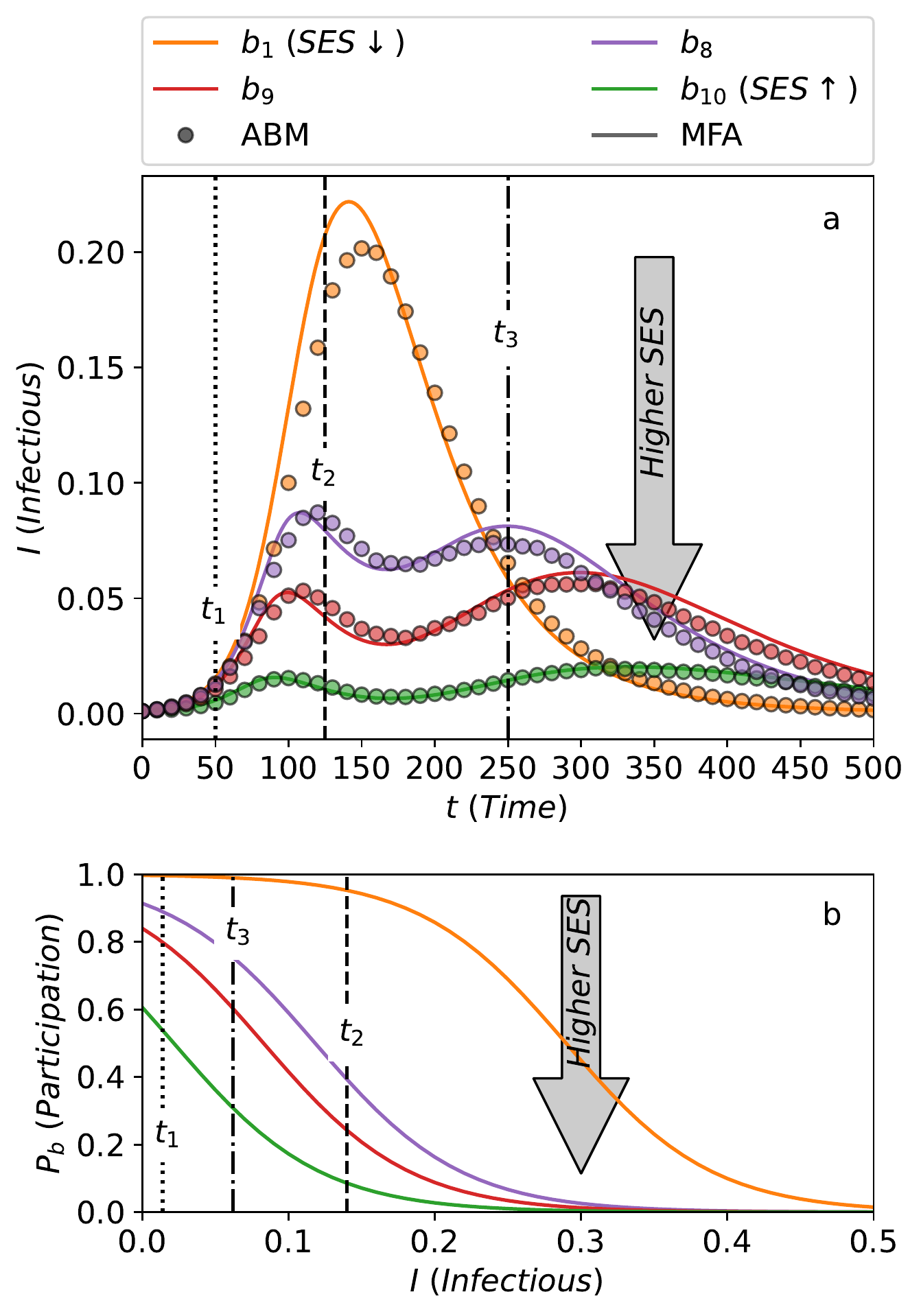}
		\caption{
		Time evolution of the spreading and decision-making dynamics for four representative blocks.
		a) Evolution of $I(t)$, the proportion of infectious agents in each block over time $t$.
		The dots and the lines respectively represent the ABM (agent-based model) and the MFA (mean-field approximation) results. \textit{Error bars for the ABM results are smaller than the marker size.}
		b) Probability of participation, $P_b$ (Eq. \ref{eq:Prob-main}) for each block $b$ as a function of the proportion of infectious population $I$.
		Color code denotes SES groups, where higher SES groups are assigned to higher indices.
		The dotted, dashed and dash-dotted vertical lines indicating the characteristic times $t_1$, $t_2$, and $t_3$ are illustrated on panel a. The lines denote their corresponding $I(t)$ values on panel b.
		}
		\label{fig:I-time-series}
	\end{center}
\end{figure}

\begin{figure}[!pt]
	%\begin{center}
		\includegraphics[width=\linewidth]{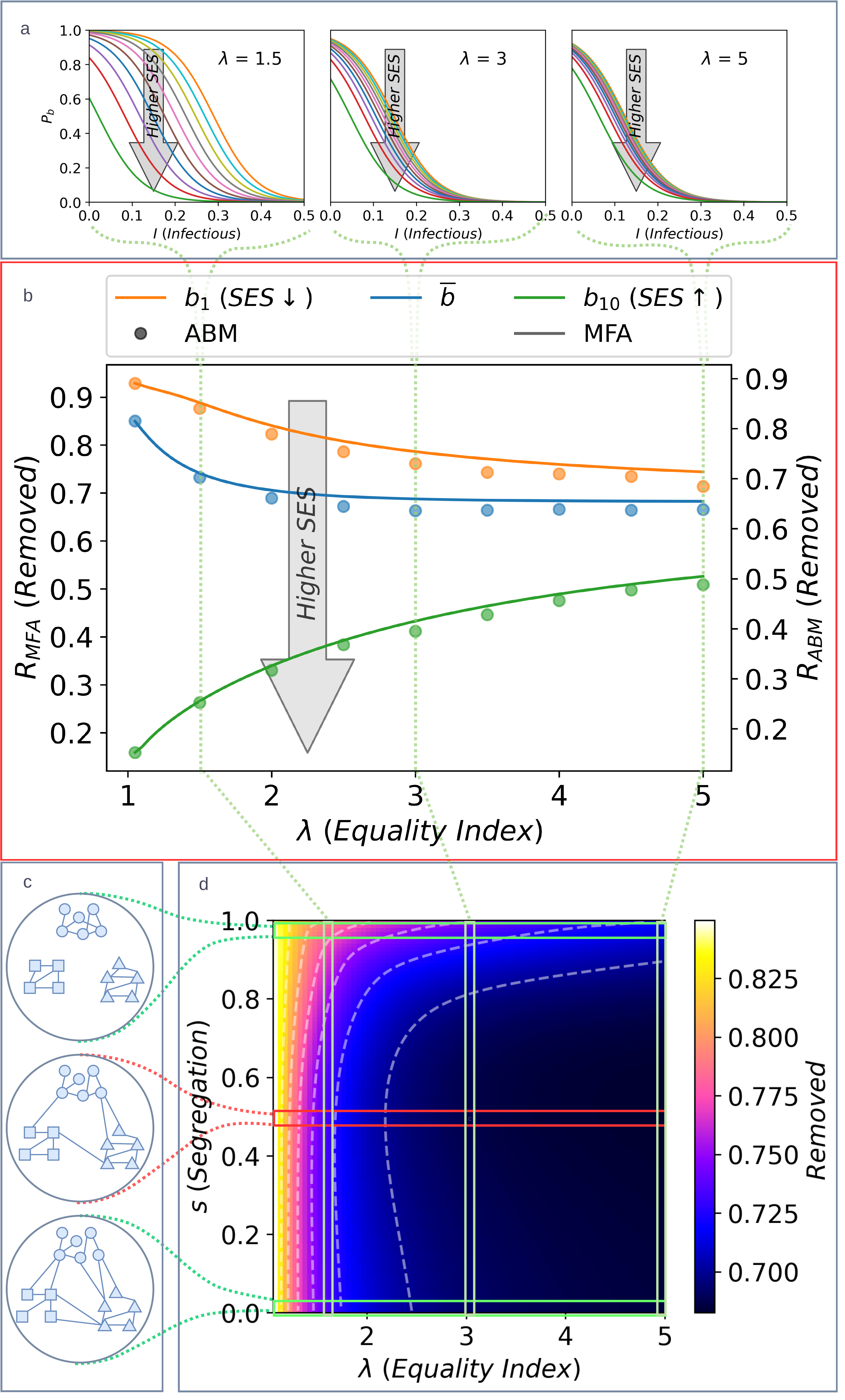}
		\caption{
		The final proportion of the removed agents, $R$ for varying values of segregation $s$ and equality $\lambda$.
		a) Probability of participation, $P_b$ (Eq. \ref{eq:Prob-main}) for each block $b$ as a function of the proportion of infectious population $I$. The arrows indicate moving toward the higher SES groups.
		Each diagram corresponds to a unique value of $\lambda$. More equal wealth distribution leads to less deviating probabilities.
		b) Final proportion of the removed agents, $R$, in each block, for varying levels of wealth equality index $\lambda$.
		Color code denotes SES groups, with higher SES being indexed by higher values. $\overline{d}$ represents the average over all SES groups.
		The dots and the lines respectively denote the ABM (agent-based model) and the MFA (mean-field approximation) results.
		The left and right y-axes respectively indicate MFA and ABM; please note the minor discrepancy in their values.
		The vertical lines connect the values of $\lambda$ to their corresponding decision-making functions.
		\textit{Error bars for the ABM results are smaller than the marker size.}
		c) Schematic network illustrations, for varying values of segregation. The number of intra-block and inter-block links respectively increase and decrease with $s$.
		d) Final proportion of the removed agents, $R$ denoted by the color-axis, for varying levels of $s$ and $\lambda$ based on the MFA.
		The vertical lines connect the values of $\lambda$ to their corresponding decision-making functions, in panel a.
		The horizontal lines connect the values of $s$ to their corresponding networks, in panel c.
		The values illustrated in the red window are also represented as the blue curve of panel b.
		}
		\label{fig:s-h-htmap}
	%\end{center}
\end{figure}

\section{Results}
We simulate the dynamics using both the agent-based model and the mean-field approximation, considering $N_b = 10$ equally populated SES blocks.
Unless stated otherwise, we use a set of baseline parameters (described in Table \ref{tab:parameters}) for all the analyses conducted in this section.
To check the robustness of the calculations we also vary the values of equality index $\lambda$, segregation $s$, and transmission reproduction rate $R_0$.
The ABM results represent an average of $100$ realizations.

\subsection*{Temporal patterns}
We study the evolution of the fraction of infectious individuals in each SES group $I_b(t)$, over time using both the agent-based model (ABM) and the mean-field approximation (ODE). The result of this analysis is illustrated in Fig. \ref{fig:I-time-series} panel a for four representative blocks.
Fig. \ref{fig:I-time-series} panel b illustrates $P_b$ the probability of participation for the selected blocks, as a function of $I(t)$.

We observe a strong agreement between the results of the two models. The slight overestimation in the mean-field (ODE) is only to be expected, as mean-field approximations generally display faster spreading dynamics. This is for the most part, due to the negligence of the network clustering in the mean-field approximation and reducing the chance of the dynamics getting trapped in closely knitted neighborhoods \cite{KeelingRohani+2011}.

Initially, at $t = t_1$ the small proportion of infectious people leads to a high probability of participation for all groups, hence the infection spreads quickly in all blocks.
As the lower SES groups have a low opportunity of quarantining themselves, they continue participating in daily life, their infection rates, therefore, increase similarly to the baseline SIR model \cite{barrat}. It peaks at $t=t_2$ and then at $t=t_3$, as there are not enough susceptible agents remaining, it drops. In other words, they transmit the disease until they achieve the state of herd immunity.

On the other hand, the higher SES groups react to the high proportion of infectious cases ($t=t_2$) and start quarantining at a higher rate, leading to a drastic decrease in their infectious rates.
As the lower SES groups achieve herd immunity and the number of active infectious cases drops ($t = t_3$), the higher SES having a \textit{false sense of safety} move out of quarantine, giving rise to a second peak.
\subsection*{Role of wealth inequality}
\label{sec:role-of-inequality}
Next, we investigate the role of wealth distribution in the epidemic dynamics.
In our model, $\lambda$ represents wealth equality in society. As illustrated in Fig. \ref{fig:s-h-htmap} panel a, $\lambda$ affects the quarantine decision making probabilities $P_b$. Increasing $\lambda$ (more equal regimes) leads to more homogeneous quarantining probabilities across the different SES blocks. 
Figure \ref{fig:s-h-htmap} panel b illustrates the final proportion of the removed agents for the lowest SES, the highest SES, and the average population. By increasing $\lambda$ and providing a more equal opportunity of quarantining, the infection prevalence (manifested in $R$) will increase for the higher SES groups. However, the lower SES groups, and most importantly, the overall population will benefit from a significant drop in their infected cases. Therefore, the society would profit from a more equal distribution of wealth and consequently a more equal affordability of quarantining. 
\subsection*{Role of social segregation}
We analyze the effect of segregation on the spreading dynamics.
As illustrated in Fig. \ref{fig:s-h-htmap} panel c, $s$ controls the probability of interaction between agents of different SES groups in the contact network. With $s=0$ (bottom) we generate the networks using a uniform probability matrix, leading to a fully homogeneous network. With $s=1$ the probability of interaction is only non-zero for agents of the same block, leading to a fully segregated network. For any value of $0<s<1$ the probability matrix is a linear combination of the two aforementioned matrices. One should note that the average number of degrees, $\langle k \rangle$ is the same for networks with different $s$ vales.

We perform the same analysis conducted in section \ref{sec:role-of-inequality},
this time also varying the level of segregation $s$. The results of this analysis are illustrated in Fig. \ref{fig:s-h-htmap} panel d, where we demonstrate the average proportion of the final removed population for different values of segregation $s$ and equality index $\lambda$. As it is evident, keeping $\lambda$ constant and increasing $s$ (moving vertically towards the top of the plot), significantly increases the final proportion of removed agents.
This result can be counter-intuitive as higher values of segregation indicate a lower interaction rate between different groups and this might imply the confinement of the disease in the segregated blocks.
However, high segregation yields several neighborhoods with a high concentration of agents having low quarantine opportunity, and hence a high infection risk.
The high proximity of these agents can give rise to an explosion of infection throughout the neighborhood, boosting the spread of the disease in society.
On the other hand, in a more uniform contact network, the quarantine performed by the high SES individuals, can also protect their low SES neighbors.

Furthermore, we conclude that the results of section \ref{sec:role-of-inequality} are robust with respect to variations in the level of segregation. For every value of $s$, the society benefits from a drop in the removed population, as we move toward the higher values of $\lambda$.
\begin{table}\centering
\begin{tabular}{ |c|l|c| } 
%\toprule
\hline
 \textbf{Parameter} & \textbf{Description} & \textbf{Value} \\ 
 \hline
%\hline
\hline
 \textit{Contact Network}  \\ 
\hline
\hline

$B$ & \# of SES groups & $10$ \\ %\arrayrulecolor{gray}\hline
$N_T$ & Total number of agents & $10^4$ \\ \hline
$N_b$ & Population of block $b$& $N_b = \frac{N_T}{B}$ \\ \hline
$\braket{k}$ & Mean degree & $20$ \\ \hline
$s$ & Segregation & $0.5$\\ %\arrayrulecolor{black}
\hline

\textit{Spreading Dynamics}   \\ 
\hline
\hline
 $R_0$ & Reproductive number & $ \frac{\mu \braket{k} }{\gamma}= 3 $ \\  %\arrayrulecolor{gray}\hline
$I_{init}$ & Initial infectious fraction & $0.001$ \\ \hline
$\gamma$ & Removal rate & $0.03$ \\ \hline
$\mu $ & Transmission rate & $ 0.0045 $ \\ \hline %\arrayrulecolor{black}\hline
\textit{Decision Making}  \\ 
\hline
\hline

$r_d$ & Fear of infection & $-10$ \\ \hline
$\beta$ & intensity of selection & $2$ \\ \hline
$\lambda$ & Indicator of equality & $1.5$\\
\hline
\end{tabular}
\caption{Model parameters and their baseline values.}
\label{tab:parameters}
\end{table}

\section{Discussion}
With this modeling framework, we are able to evaluate the impact of two important dimensions of structural inequality, income inequality, and social segregation, on the health disparities in society. One key finding is that the level of income inequality in society matters in how the infection cases are distributed across social groups. When the income gap is high ––because more low-income people will be obliged to go out to seek their livelihood–– the infection rate will be higher than in a more equal society, where fewer people will have to go out to go to work despite the infection risk. Second, we find that higher segregation significantly intensifies the infection and subsequently mortality rates.
It might seem that high segregation and low interaction among the groups, restricts the possibility of transmission across different SES groups, resulting in a lower infection rate.
However, a high concentration of low SES individuals unable to quarantine can bring about a burst of infection in society.
On the other hand, in a more uniform society, the low SES will be partially protected by the quarantine of their higher SES neighbors.
Our computational model also predicts the appearance of the second peak in infection rates, as a result of the interplay between the economic inequality and decision-making procedure of the population. The low SES population gets infected until it reaches the point of herd immunity, the high SES on the other hand can afford to quarantine. However, after a drop in the low SES infectious cases, the high SES perceive the danger of infection to be low and move out of quarantine. This subsequently leads to a second infection peak.

Methodologically, this work is among the first papers to combine spreading dynamics on networks and game theory to model dynamics of epidemic inequalities during a public health crisis. This computational framework allows us to compare different scenarios with respect to segregation and wealth inequality in a systematic manner. Moreover, we investigate the outcome of the simulations with the analytical approximation, enabling us to interpret the results in a reliable and robust way.

This modeling study is best being evaluated on the onset of a new pandemic when vaccination solutions are not in place. In the future, it would be interesting to study how the delay in vaccinating lower-income classes can exacerbate health inequality and the time to reach immunity. In this paper, we assumed that each SES group has the same network density. In reality, lower-income groups often live in overpopulated housing estates in neighborhoods that suffer from a shortage of critical care physicians and medical supplies and work under conditions that do not allow 6-foot distancing \cite{Yancy:2020aa}. Nevertheless, the results of this model assumption can serve as the minimal condition of infection spread among the groups. One direction for future research would be to include the dynamic of income inequality as a by-product of quarantining decisions. Finally, empirical data with information about the socioeconomic status of individuals, the willingness to participate in daily activities, and combined with spatial social networks would be beneficial to calibrate and validate the computational model. 
In a broader context, health inequalities can enforce other forms of inequalities and discrimination that should be tackled in a systematic and data-driven manner. This computational model can be served as a tool to predict the severity of the infection disparity within a population and to examine various intervention scenarios to avoid health disparities in society. Policymakers need to acknowledge the different behavioral implications that working and living conditions generate to formulate adequate policy actions during health crises.

\section*{Data and code availability}
All the simulations, analyses and illustrations have been conducted via the \texttt{unequal-spread} package (written in Python), developed by Sina Sajjadi and Pourya Toranj Simin.
This package is available under GPLv3, at \url{https://github.com/Sepante/unequal-spread}.

\section*{Acknowledgments}
Authors thank Zahra Rezazade for her help in designing Fig. \ref{fig:model-illus}.

%\bibliography{cas-refs.bib}

%% Loading bibliography style file
%\bibliographystyle{model1-num-names}
%\biboptions{sort}
% Loading bibliography database
%\bibliography{cas-refs}
%\printbibliography
%\bibliographystyle{cas-model2-names}

%\bibliographystyle{ieeetr}

%\begin{comment}

%apsrev4-2.bst 2019-01-14 (MD) hand-edited version of apsrev4-1.bst
%Control: key (0)
%Control: author (8) initials jnrlst
%Control: editor formatted (1) identically to author
%Control: production of article title (0) allowed
%Control: page (0) single
%Control: year (1) truncated
%Control: production of eprint (0) enabled
%

%\end{comment}

\clearpage
\setcounter{section}{0}
\setcounter{subsection}{0}
%\setcounter{table}{0}
%\setcounter{figure}{0}
%\setcounter{equation}{0}

%\setcounter{page}{1}

%\appendix 
%\renewcommand\thefigure{\Roman{figure}}
\renewcommand\thesection{S\arabic{section}}

\begin{center}
\textbf{\large Supplementary Information}
\end{center}
\setcounter{equation}{0}
\setcounter{figure}{0}
\setcounter{table}{0}
\setcounter{page}{1}
\makeatletter
\renewcommand{\theequation}{S\arabic{equation}}
\renewcommand{\thefigure}{S\arabic{figure}}
\renewcommand{\bibnumfmt}[1]{[S#1]}
\renewcommand{\citenumfont}[1]{S#1}

\section{Agent-based simulation details}
In this section, we provide the details of the agent-based simulation, necessary for the reproduction of our results.
In this model each agent can be in one of the three different states: \enquote{SIR} (Susceptible-Infected-Removed).
Agents are also classified into SES blocks, affecting their decision on whether or not to quarantine at each time step.

The Initial state, dynamics, and the final state of the simulation are described in the following sections.

\subsection{Initial State}
A proportion of size $I_{init}$ within each SES block is set to state $I$, the rest of the population is in state $S$.

\subsection{Dynamics}
The infection simulations are conducted using the rejection-based modeling \cite{rejection-based}. In this method, all possible actions (infection, removal, or quarantine) will be proposed at each time step and they will be either accepted or rejected, based on their corresponding probability.
The simulation steps consist of the following:
\begin{enumerate}
    \item \textit{Quarantine phase:} Every agent decides to quarantine, with the probability $1-P_b(I)$, based on its SES block membership, $b$ and the total infectious fraction of the population, $I(t)$ at the time.
	\item \textit{Infection phase:} Every non-quarantining infectious agent, turns its non-quarantining susceptible neighbors to infectious with the infection probability $\mu \Delta t$, where $\mu$ is the transmission rate and $\Delta t$ is the size of a time step.
	\item \textit{Removal phase:} Every infectious agent gets removed with the removal probability $\gamma \Delta t$, where $\gamma$ is the rate of removal \cite{barrat}.
\end{enumerate}
As we employ a parallel updating algorithm, the transitions in agents' health status only take effect after the end of each time step.
Therefore, an agent turning to infectious at time step $t$, can only transmit the infection and/or get removed, in time steps $t'$, where $t'>t$.

\subsection{Final State}
The simulation runs until it reaches the stationary state, i.e., until all of the agents are in one of the states, $S$ or $R$.

\section{Mean-field approximation}
\label{sec:mean-field-derivation}
In this section, we derive the mean-field approximation for our agent-based model. For this purpose, we first define the concept of \enquote{compartment} as the set of agents with the same SES and health (S-I-R) status. Secondly, we take into account the following assumptions:

\begin{enumerate}

\item Homogeneous Mixing: Agents within each compartment are identical and equally exposed to the rest of the population.
\item Determinism: 
The stochastic fluctuations of the system can be neglected as they are considered inconsequential on large scales.
\item Continuity: The progress of the dynamics can be approximated by a continuous time dimension instead of discrete time steps.
\end{enumerate}
As a result of the first assumption, the state of the system can be represented by the size of the compartments (fractions of $S$, $I$, and $R$ agents within each SES group). By incorporating the second and third assumptions, the dynamics can be described by a set of exchange rates between the compartments in the form of differential equations.

The transitions in the model consist of the following:

\begin{equation}\label{transition:transmission}
    S + I \rightarrow I + I
\end{equation}

\begin{equation}\label{transition:removal}
    I \rightarrow R
\end{equation}

To obtain the transition rates between the compartments, we formulate Eq. \ref{eq:omega}, where $\widetilde{\Omega_i}$ is the probability of the transmission (Eq. \ref{transition:transmission}) \underline{not} to occur for an agent $i$ of block $b$, during a time period of $\Delta t$. The first term on the right hand side denotes the case in which agent $i$ is under quarantine, and the second term represents the situation where agent $i$ is not in quarantine ($P_b$) but does not get infected ($Q_i$).

\begin{equation}\label{eq:omega}
    \widetilde{\Omega_i} = (1 - P_b) + P_b Q_i
\end{equation}

To derive $Q_i$, we focus on the probability of transmission from a specific agent $j$ in block $b'$. For this transmission to occur,
agent $j$ should be both infectious and out of quarantine.
Given that this condition holds for a period of $\Delta t$, the probability of transmission would be $\mu \Delta t$, as $\mu$ is the infection probability in one time-step in the discrete model.
Therefore, $\Omega_{j \rightarrow i}$ the overall probability of a susceptible agent $i$ which is out of quarantine, getting infected by agent $j$ from block $b'$ is as formulated in Eq. \ref{eq:1}.

\begin{equation}\label{eq:1}
    \Omega_{j \rightarrow i} = P_{b'} \frac{I_{b'}}{N_{b'}} (\mu \Delta t)
\end{equation}

Where $I_{b'}$ is the number of infected agents in block $b'$ and $N_{b'}$ is the population of block $b'$; therefore due to the homogeneous mixing assumption, $\frac{I_{b'}}{N_{b'}}$ is the probability that agent $j$ from block $b'$ is infected.

The probability $Q_i$ that the not-quarantined agent $i$ from block $b$, is not getting infected by any of its neighbors is obtained in Eq. \ref{eq:2}.

\begin{equation}\label{eq:2}
\begin{split}
&
Q_i
=\prod^{k_{i}}_{j=1}
( 1-\Omega_{j \rightarrow i} )
=\prod^{k_{i}}_{j=1}
(1-P_{b'}\frac{I_{b'}}{N_{b'}}\mu \Delta t)
\\
&
=\prod^{B}_{b'=1} \prod^{\text{\{Neighbours of i in $b'$\}}}_{j}
(1-P_{b'}\frac{I_{b'}}{N_{b'}}\mu \Delta t)
\\
&
= \prod^{B}_{b'=1} (1-P_{b'}\frac{I_{b'}}{N_{b'}} \mu \Delta t)^{k_{i, b'}}
\end{split}
\end{equation}

Where $k_{i, b'}$ is the number of $i$'s neighbors within block $b'$. As $\phi_{b,b'}$ is the share of neighbors of $i \in b'$ which are from block $b$.
 then $k_{i,b'} \approx \phi_{b,b'} k_i$ where $k_i \approx \braket{k}$ is the total number of $i$'s neighbors.
(see section \ref{sec:share-of-connection})
We can therefore, approximate Eq. \ref{eq:2} as in Eq. \ref{eq:3}.
\begin{equation}\label{temp}
\end{equation}
\begin{equation}\label{eq:3}
\begin{split}
&
Q_i = \prod^{B}_{b'=1}(1-P_{b'}\frac{I_{b'}}{N_{b"}}\mu \Delta t)^{k_{i, b'}}
\\
&
\approx \prod^{B}_{b'=1}(1-P_{b'}\frac{I_{b'}}{N_{b'}} \mu \Delta t)^{\phi_{b,b'} \braket{k}}
\\
&
\approx \prod^{B}_{b'=1} (1-\phi_{b,b'} \braket{k} P_{b'}\frac{I_{b'}}{N_{b'}}\mu \Delta t)
\\
&
\approx \prod^{B}_{b'=1} (1-\rho_{b,b'} P_{b'}I_{b'}\mu \Delta t)
= \prod^{B}_{b'=1}(1-\alpha_{b'} \Delta t)
\end{split}
\end{equation}
Where we have defined $\alpha_{b'} \equiv \rho_{b,b'} P_{b'}I_{b'}\mu$, and by rephrasing Eq. \ref{eq:3} we have Eq. \ref{eq:4}.

\begin{equation}\label{eq:4}
    \begin{split}
&
    \prod^{B}_{b'=1}
(1-\alpha_{b'} \Delta t)  = 
\exp{[\ln{(\prod^{B}_{b'=1}(1-\alpha_{b'} \Delta t))}]}
\\
&
=
e^{\sum^{B}_{b'=1}\ln(1-\alpha_{b'} \Delta t)}\\
\end{split}
\end{equation}
As $\Delta t \ll 1 $ we can approximate $\ln(1-\alpha_{b'} \Delta t) \approx -\alpha_{b'} \Delta t$ obtaining Eq. \ref{eq:5}.
\begin{equation}\label{eq:5}
\begin{split}
    Q_i \approx  &\prod^{B}_{b'=1}
(1-\alpha_{b'} \Delta t)  \approx
e^{-\sum^{B}_{b'=1}\alpha_{b'} \Delta t}
\\
&
= 1 - \sum^{B}_{b'=1}\alpha_{b'} \Delta t + O(\Delta t^2) + ...\\
& \approx 1 - \sum^{B}_{b'=1}\alpha_{b'} \Delta t = 1 - \sum^{B}_{b'=1}\rho_{b,b'} P_{b'} I_{b'} \mu \Delta t\\ 
\end{split}
\end{equation}
Knowing $Q_i$, we derive $\Omega_i$, the probability that agent $i$ from block $b$ gets infected by any of its neighbours in Eq. \ref{eq:6}.
\begin{equation}\label{eq:6}
    \Omega_i = 1- \left(1-P_{b}+ P_{b} Q_i\right)
\end{equation}
By substituting $Q_i$ in Eq. \ref{eq:6} we obtain Eq. \ref{eq:7}.
\begin{equation}\label{eq:7}
\begin{split}
&
\Omega_i\approx 
1- \left(1 - P_{b}+P_{b}(1-\sum^{B}_{b'=1} \rho_{b,b'}  P_{b'} I_{b'} \mu \Delta t)\right) 
\\
&
= \mu P_{b} \sum^{B}_{b'=1} P_{b'} I_{b'} \rho_{b,b'} \Delta t
\end{split}
\end{equation}

The expected value of the number of infected agents after a time duration of size $\Delta t$ and hence the change in the number of susceptible agents $S_b$ in block $b$ can be derived as in Eq. \ref{eq:8}.
\begin{equation}\label{eq:8}
    \begin{split}
    &
        \Delta S_b = - \sum_{i \in S_b} \Omega_i \approx - \sum_{i \in S_b} \mu P_{b} \sum^{B}_{b'=1} P_{b'} I_{b'} \rho_{b,b'} \Delta t 
        \\
        &
        =-\mu \left( \sum_{i \in S_b} \right)\left( 
         P_{b} \sum^{B}_{b'=1} P_{b'} I_{b'} \rho_{b,b'} \Delta t 
        \right)
        \\
        &
        =-\mu S_b P_{b} \sum^{B}_{b'=1} P_{b'} I_{b'} \rho_{b,b'} \Delta t 
    \end{split}
\end{equation}
By differentiating $S_b$ with respect to time we arrive at Eq. \ref{eq:9}.
\begin{equation}\label{eq:9}
    \dfrac{d S_b(t)}{d t} = -\mu S_{b}(t) P_{b} \sum^{B}_{b'=1} P_{b'} I_{b'} \rho_{b,b'}
\end{equation}

On the other hand
due to the transition in Eq. \ref{transition:removal},
the infected agents will get removed with a probability of $\gamma \Delta t$. Therefore, $\Delta R_b$ the number of infectious agents in block $b$ that will get removed after a time duration of size $\Delta t$ can be expressed in Eq. \ref{eq:10}.
\begin{equation}\label{eq:10}
    \Delta R_b = I_b \gamma \Delta t
\end{equation}
Resulting in Eq. \ref{eq:11}.
\begin{equation}\label{eq:11}
    \dfrac{d R_b(t)}{d t} = \gamma I_b(t)
\end{equation}

As a result of the conservation of the population of each block we also have Eq. \ref{eq:12}.
\begin{equation}\label{eq:12}
    S_b(t)+I_b(t)+R_b(t) = N_b
\end{equation}
By putting the equations \ref{eq:9}, \ref{eq:11} and \ref{eq:12} together, we arrive at the set of ordinary differential equations describing the dynamics of the system in Eq. \ref{eq:dynamics}.

\begin{equation}\label{eq:dynamics}
\begin{split}
	&
	  \dfrac{d S_b(t)}{d t} = -\mu  S_{b}(t) P_{b} \sum^{B}_{b'=1} P_{b'} I_{b'} \rho_{b,b'}
	\\
	&
	\dfrac{d R_b(t)}{d t} = \gamma I_b(t)
	\\
	&
   S_b(t)+I_b(t)+R_b(t) = N_b
	\end{split}
\end{equation}

In the main text, Fig. \ref{fig:I-time-series} illustrates the mean-field value of the infectious compartment size over time, ($I_b(t)$ in Eq. \ref{eq:dynamics}) as solid lines.
Fig. \ref{fig:s-h-htmap} panel b illustrates the final ($t \rightarrow \infty$) mean-field value of the removed compartment size ($\lim_{t \to \infty} R(t)$ in Eq. \ref{eq:dynamics}) as solid lines in panel b.
Fig. \ref{fig:s-h-htmap} panel d, Fig. \ref{fig:R0-h-htmap} and Fig. \ref{fig:wd-l-htmap} illustrate the same parameter as the color code.

\section{Segregation}
To model segregated networks, we use stochastic block models (SBMs). SBMs are capable of demonstrating a controllable of modularity \cite{newman2006modularity} i.e., higher intra-group and lower inter-group connections, making them a good candidate for modeling community structures.
In the stochastic block model, every two agents, $i$ and $j$ respectively belonging to blocks $a$ and $b$, are connected with probability $\rho_{a,b}$, where $\boldsymbol{ \rho }$ is known as the probability matrix.
Here we consider the \enquote{blocks} to be identical to socioeconomic groups.

First, we construct a fully segregated $\boldsymbol{ \rho^{(\text{hg})} }$ ($\text{hg}$ for heterogeneous) and a fully homogeneous matrix $\boldsymbol{ \rho^{(\text{hm})} }$ ( ($\text{hm}$ for homogeneous)) .
In the former case, only links within blocks are allowed. Therefore, to maintain a uniform mean degree $\braket{k}$ across the blocks, we set $\rho^{(\text{hg})}_{a,b} = \delta_{a,b} \frac{ \braket{k} } {N_a}$ with ${N_a}$ indicating the number of nodes within block $a$. In the latter case, agents do not discriminate based on block memberships and therefore $\rho^{(\text{hm})}_{a,b} = \frac{ \braket{k} } {N_T} $ with ${N_T}$ indicating the total number of the nodes.

To construct a probability matrix with an arbitrary level of segregation we define $\boldsymbol{ \rho^{(s)} } = s \times \boldsymbol{ \rho^{(\text{hg})} } + (1-s) \times \boldsymbol{ \rho^{(\text{hm})} }$
as a linear combination of the aforementioned matrices, where $s$ indicates the intensity of segregation in the matrix.

\begin{figure}[ht]
	\begin{center}
		\includegraphics[width=\linewidth]{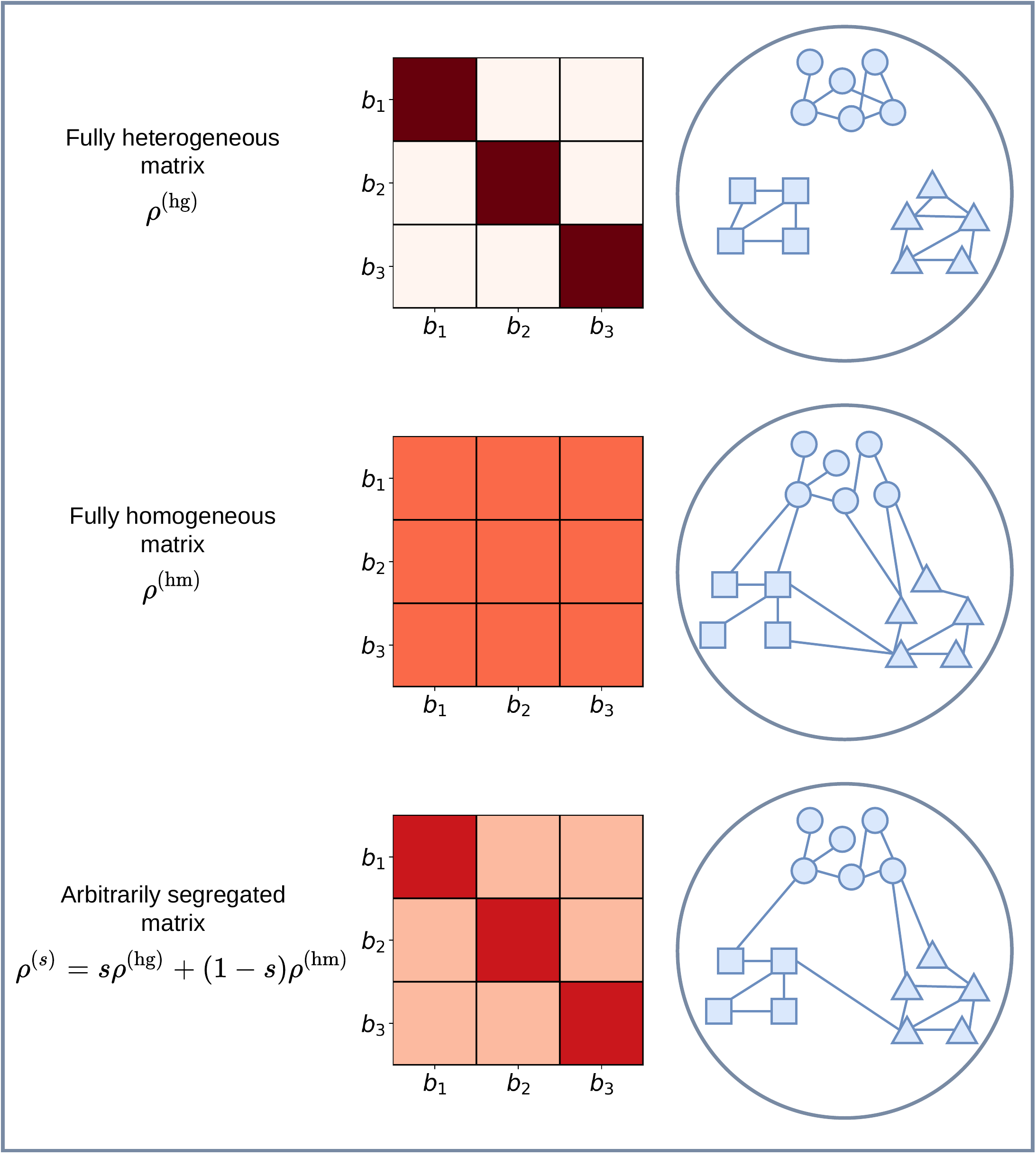}
		\caption{Segregation in the contact network: The value of segregation controls the interaction between individuals of different SES groups. By changing the value of $s$, we move from a fully homogeneous society to a fully segregated regime, so the on-diameter and off-diameter values of the probability matrix respectively increase and decrease. This variation is also reflected in the toy network illustrations. In the toy illustrations, shapes denote socioeconomic blocks. By increasing the value of $s$ the number of intra-block and inter-block links respectively increase and decrease.
		}
		\label{fig:s-network}
	\end{center}
\end{figure}

Using $ \boldsymbol{ \rho^{(s)} }$ we generate networks with desired segregation levels, as illustrated in Fig. \ref{fig:s-network}.

\subsection{Share of Connections}\label{sec:share-of-connection}
To calculate  $\phi_{a,b}$, the share of neighbors from block $b$ for agent $i \in a$,
we divide the average number of $i$'s neighbors in block $b$ by the average number of $i$'s total neighbors. Therefore, we obtain $\phi_{a,b} = \frac{\rho_{a,b} N_b}{ \braket{k} }$.\
It should be noted that in contrast to $\boldsymbol{ \rho }$, $\boldsymbol{ \phi }$ is not symmetric and is invariant under the population size $N_T$.

\section{Economic Inequality}
\label{sec:economic-inequality-appendix}

The process of decision-making is governed by an economic game, where agents decide whether to quarantine themselves or to participate in society. Going to work exposes the agent to the infection, resulting in a \textit{psychological fear of infection punishment}. The option of quarantining, in contrast, leads to \textit{income loss}. The effect of income loss is higher for individuals from lower socio-economic classes, as they will be less likely to afford to lose their income and quarantine themselves. Hence, we hypothesize the punishment of income-loss for each socio-economic class to be inversely proportional to its share of wealth.

We assume wealth $w$ is distributed in the society, by the Pareto distribution $p(w) =  \frac{\lambda w_m^\lambda}{w^{\lambda+1}}$. The parameter $\lambda$ controls the level of equality, where $\lambda \rightarrow 0$ and $\lambda \rightarrow \infty$ respectively represent the ultimate inequality (Dirac delta distribution of wealth) and equality (Uniform distribution of wealth). Distribution of wealth in turn controls the fear of income-loss ($r_b = -\frac{1}{w_b}$).
Knowing the wealth distribution, we can calculate the share of wealth for different SES groups. Let $F$ be the cumulative portion of the population from poorest to richest, and $L$ be the cumulative share of wealth; then, for Pareto distribution, we have:
\begin{equation}
    L(F) = 1- (1-F)^{\frac{\lambda -1}{\lambda}}
~~~~\end{equation}

\begin{figure}[ht]
	\begin{center}
		\includegraphics[width=0.8\linewidth]{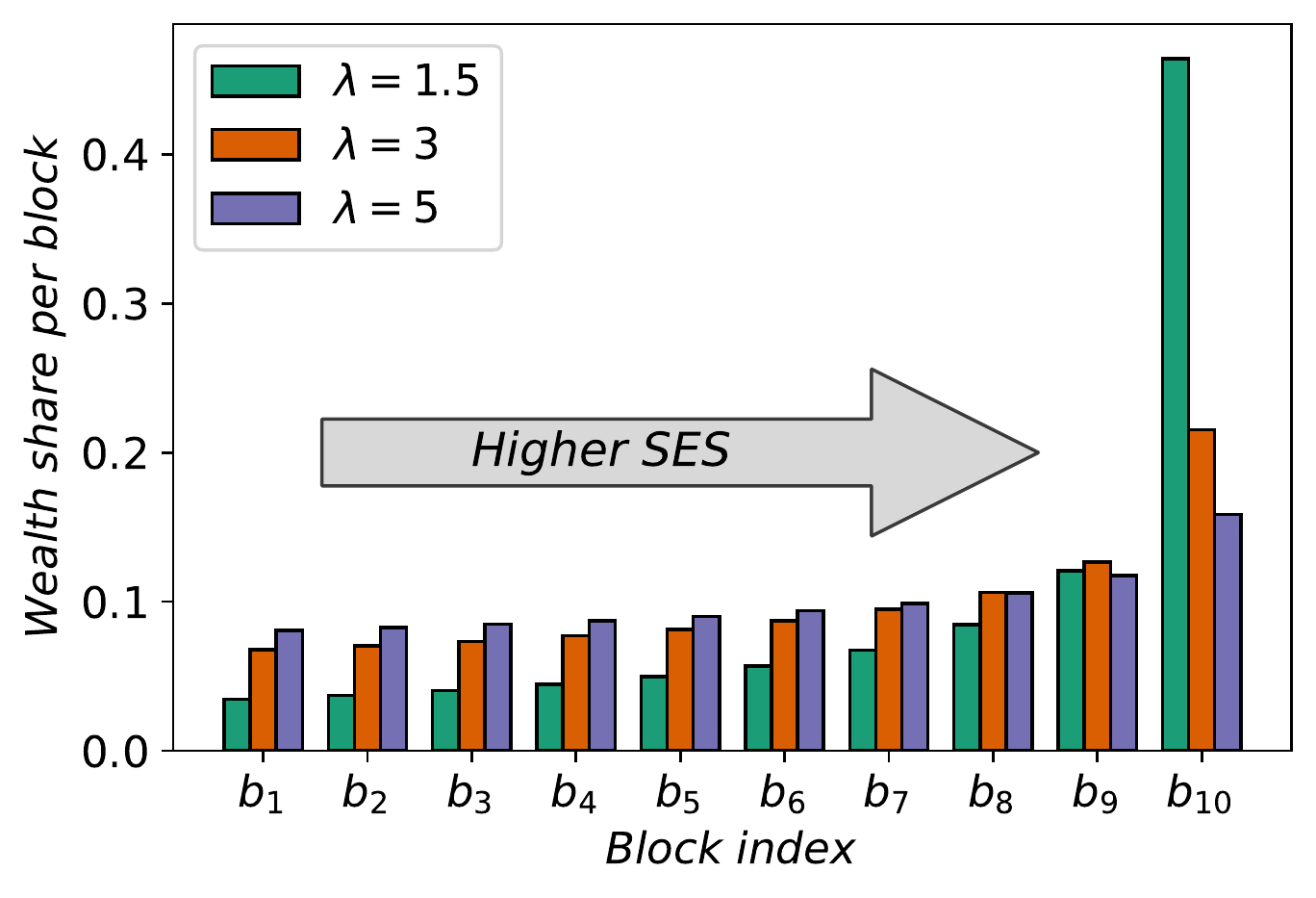}
		\caption{Distribution of the wealth among $10$ blocks, via the pareto distribution with exponents $\lambda = 1.5, 3, 5$.
		The arrow indicates moving from the lower to higher SES groups.
		}
		\label{fig:pareto}
	\end{center}
\end{figure}

This equation indicates the proportion of wealth owned by the $F$ fraction of the population.
In Fig. \ref{fig:pareto} we illustrate the share of wealth for each block of the population, for different values of $\lambda$.

We further assume the fear of income-loss $r_b$ to be proportional to the inverse of each SES group's wealth:
\begin{equation}
   r_b = -\frac{1}{w_b}
\end{equation}

We thus, hypothesize the probability $P_b$ of \underline{not} quarantining for agents of block $b$ to be:

\begin{equation}\label{eq:Prob}
P_b(t) = \frac{1}{e^{\beta \left( r_b - r_d I(t)\right)}  + 1 }
\end{equation}

\section{Robustness Check}
In this section, we vary the baseline parameter values to further check the robustness of our analysis.

\subsection{Transmission Parameters}

By dividing the first two equations in Eq. \ref{eq:dynamics} by $\gamma$, we will obtain Eq. \ref{eq:dynamics-by-gamma}.

\begin{equation}\label{eq:dynamics-by-gamma}
\begin{split}
	&
	  \dfrac{d S_b(t)}{\gamma d t} = -\frac{\mu}{\gamma}  S_{b}(t) P_{b} \sum^{B}_{b'=1} P_{b'} I_{b'} \rho_{b,b'}
	\\
	&
	\dfrac{d R_b(t)}{\gamma d t} =  I_b(t)
	\\
	&
   S_b(t)+I_b(t)+R_b(t) = N_b
	\end{split}
\end{equation}

We can scale the time dimension $dt \rightarrow dt' = \gamma dt$. By doing so, the dynamical equations would evolve at a different time scale, however, the stationary state (final removed population) would be invariant.
In addition, considering $\phi_{a,b} = \frac{\rho_{a,b} N_b}{ \braket{k} }$ (see section \ref{sec:share-of-connection} ) and replacing $\rho$ we will have Eq. \ref{eq:dynamics-rescaled}.

\begin{equation}\label{eq:dynamics-rescaled}
\begin{split}
	&
	  \dfrac{d S_b(t)}{ d t'} = -\frac{\mu \braket{k}}{\gamma} \frac{1}{N_b}  S_{b}(t) P_{b} \sum^{B}_{b'=1} P_{b'} I_{b'} \phi_{b,b'}
	\\
	&
	\dfrac{d R_b(t)}{ d t'} =  I_b(t)
	\\
	&
   S_b(t)+I_b(t)+R_b(t) = N_b
	\end{split}
\end{equation}

As the parameters $\mu$, $\gamma$ and $\braket{k}$ only appear in the form $\frac{\mu \braket{k}}{\gamma}$, we can introduce $R_0 = \frac{\mu \braket{k}}{\gamma}$ so that all are incorporated in the parameter $R_0$. By varying $R_0$ we inspect the robustness of the results with respect to all the three aforementioned parameters.

\begin{figure}[ht]
	\begin{center}
		\includegraphics[width=\linewidth]{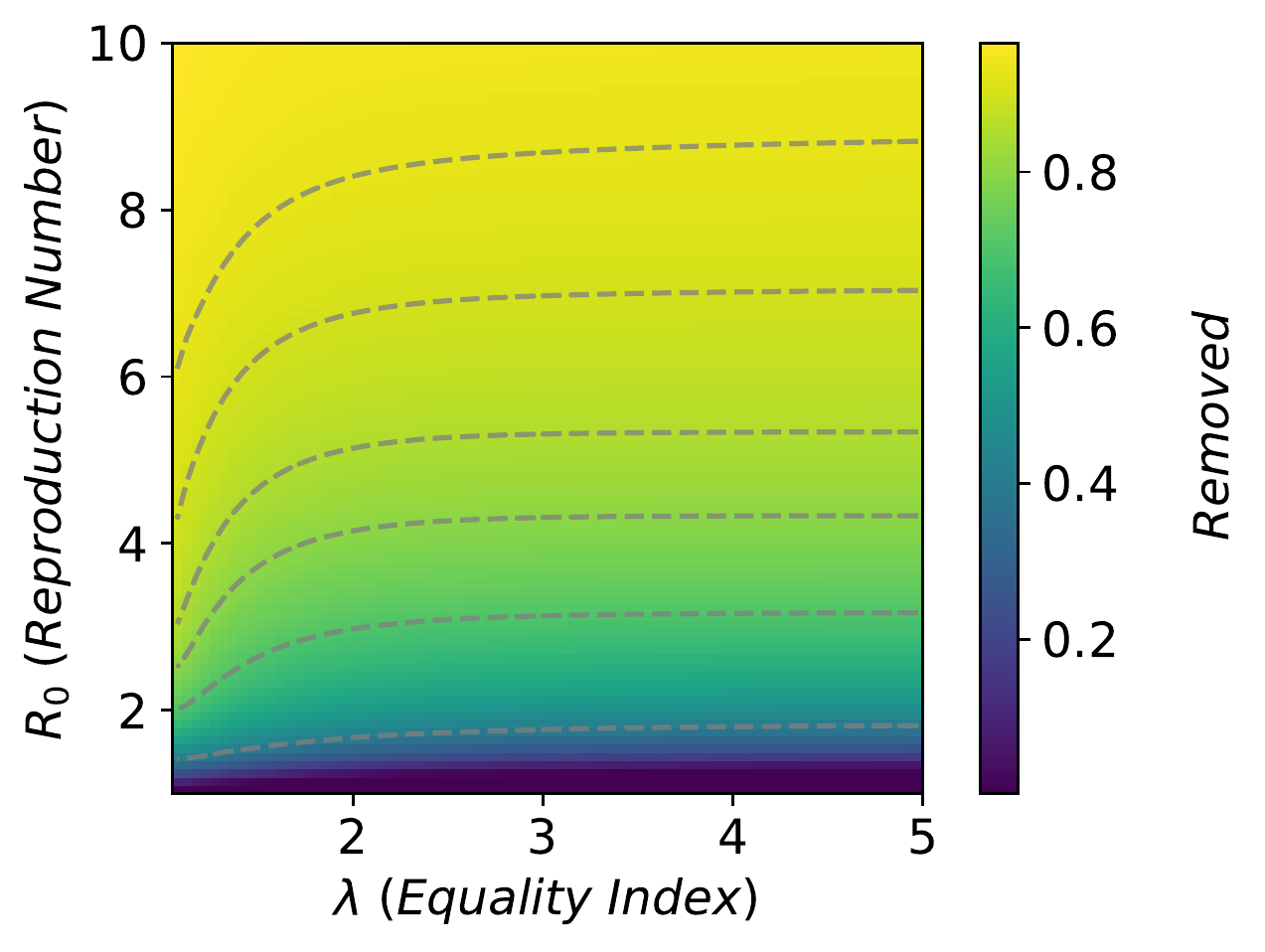}
		\caption{Final proportion of the removed agents, $R$ in the society for varying levels of transmission rate and equality. x-axis is the parameter $\lambda$, indicator of wealth equality. y-axis is the reproduction number of infection $R_0$, controlling the transmission rate of the disease.
		Color-axis represents the proportion of the removed agents $R$ at the end of the dynamics. The results are based on the mean-field approximation.
		Dashed lines are guides for the eye.
		}
		\label{fig:R0-h-htmap}
	\end{center}
\end{figure}

The results of this analysis are illustrated in Fig. \ref{fig:R0-h-htmap}. Unsurprisingly, increasing $R_0$ (moving upward) increases the final removed proportion of the population.
More importantly, increasing $\lambda$ (moving horizontally to the right) lowers the prevalence of the infection, for all $R_0$ values. Further supporting our results with respect to the $\lambda$ equality index in section \ref{sec:role-of-inequality}.

\subsection{Decision Making Parameter}
\begin{figure}[ht]
	\begin{center}
		\includegraphics[width=\linewidth]{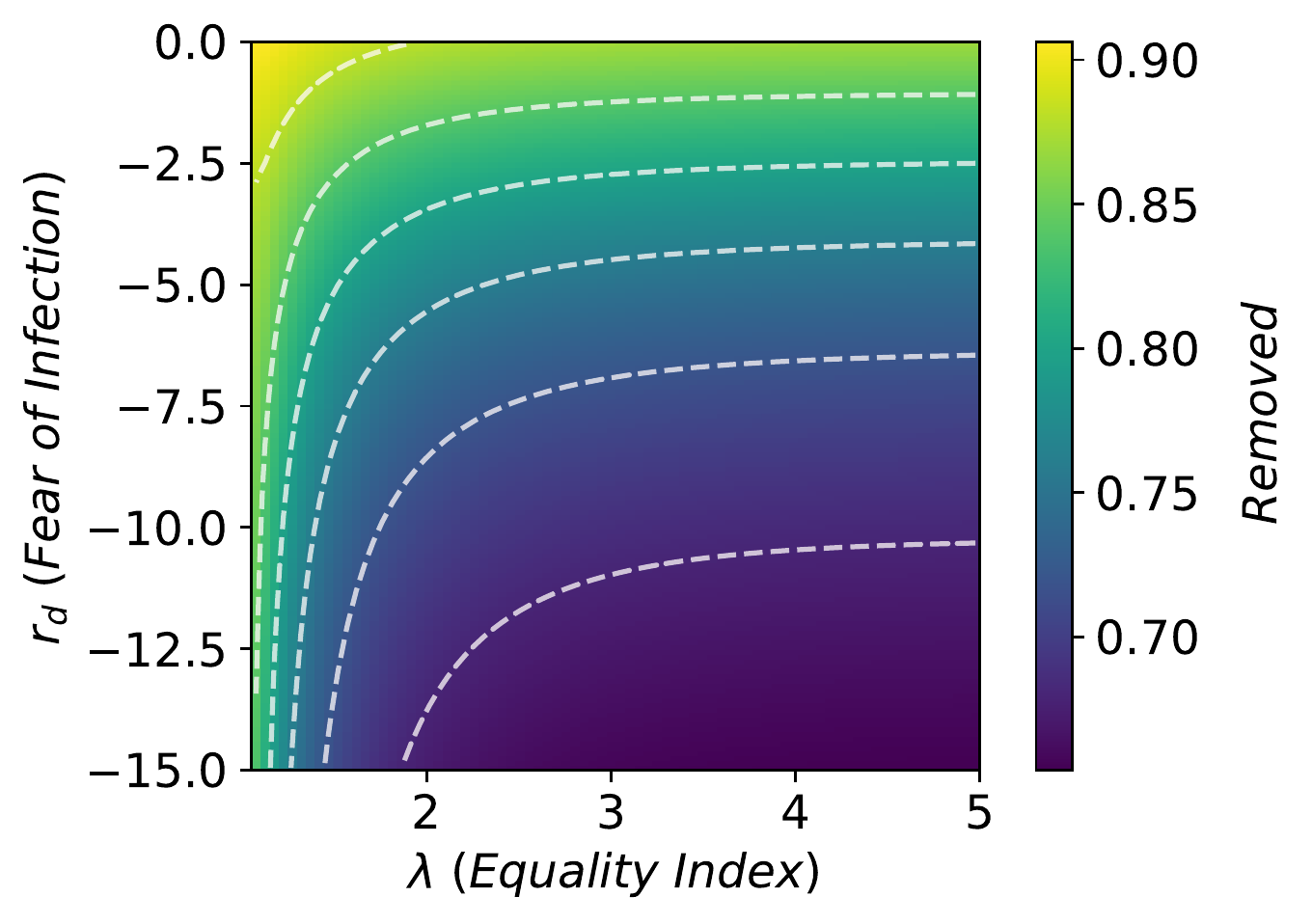}
		\caption{Final proportion of the removed agents, $R$ in the society for varying levels of fear of infection and equality. x-axis is the parameter $\lambda$, indicator of wealth equality. y-axis is the parameter $r_d$, indicator of the fear of infection.
		Color-axis represents the proportion of the removed agents $R$ at the end of the dynamics. The results are based on the mean-field approximation.
		Dashed lines are guides for the eye.
		}
		\label{fig:wd-l-htmap}
	\end{center}
\end{figure}

In this section, we analyze the robustness of our results with respect to the decision-making parameters, $r_d$..
In Fig. \ref{fig:wd-l-htmap} we have illustrated the results of this analysis.
As expected, by intensifying the fear of the infection (lower negative values), the population will act more vigilantly and therefore the final removed population would be smaller.
More importantly, increasing $\lambda$ (moving horizontally to the right) lowers the prevalence of the infection, for all $r_d$ values. Further supporting our results with respect to the $\lambda$ equality index in section \ref{sec:role-of-inequality}.

%

%\vskip3pt

\end{document}